# Transfer Learning Enhanced Generative Adversarial Networks for Multi-Channel MRI Reconstruction

Jun Lv [a], Guangyuan Li [a], Xiangrong Tong [a], Weibo Chen [b], Jiahao Huang [c], Chengyan Wang [d, *], and Guang Yang [e, f, *]

[a.] School of Computer and Control Engineering, Yantai University, Yantai, China;
[b.] Philips Healthcare, Shanghai, China;
[c.] School of Optics and Photonics, Beijing Institute of Technology, Beijing, China;
[d.] Human Phenome Institute, Fudan University, Shanghai, China;
[e.] Cardiovascular Research Centre, Royal Brompton Hospital, London, SW3 6NP, U.K.;
[f.] National Heart and Lung Institute, Imperial College London, London, SW7 2AZ, U.K.

[*] Correspondence authors: C Wang (wangcy@fudan.edu.cn) and G Yang (g.yang@imperial.ac.uk)

**Highlights**

- Improving the generalizability of image reconstruction networks based on small training samples is urgently needed.
- We developed transfer learning of a pre-trained PI-GAN model for under-sampled MRI reconstruction that can be used for the following scenarios:
    (1) Transfer learning of the pre-trained PI-GAN model for a private clinical brain test dataset with tumors.
    (2) Transfer learning of the pre-trained PI-GAN model for public knee and private liver test datasets.
    (3) Transfer learning of the pre-trained PI-GAN model (4× acceleration) for a test brain dataset with 2× and 6× accelerations.
- Using transfer learning enhances the performance of multi-channel MRI reconstruction with parallel imaging and generative adversarial networks.

**Abstract**

Deep learning based generative adversarial networks (GAN) can effectively perform image reconstruction with under-sampled MR data. In general, a large number of training samples are required to improve the reconstruction performance of a certain model. However, in real clinical applications, it is difficult to obtain tens of thousands of raw patient data to train the model since saving *k*-space data is not in the routine clinical flow. Therefore, enhancing the generalizability of a network based on small samples is urgently needed. In this study, three novel applications were explored based on parallel imaging combined with the GAN model (PI-GAN) and transfer learning. The model was pre-trained with public Calgary brain images and then fine-tuned for use in (1) patients with tumors in our center; (2) different anatomies, including knee and liver; (3) different *k*-space sampling masks with acceleration factors (AFs) of 2 and 6. As for the brain tumor dataset, the transfer learning results could remove the artifacts found in PI-GAN and yield smoother brain edges. The transfer learning results for the knee and liver were superior to those of the PI-GAN model trained with its own dataset using a smaller number of training cases. However, the learning procedure converged more slowly in the knee datasets compared to the learning in the brain tumor datasets. The reconstruction performance was improved by transfer learning both in the models with AFs of 2 and 6. Of these two models, the one with AF=2 showed better results. The results also showed that transfer learning with the pre-trained model could solve the problem of inconsistency between the training and test datasets and facilitate generalization to unseen data.

# 1. Introduction

Magnetic Resonance Imaging (MRI) is widely used for the diagnosis of diseases due to superior soft-tissue contrast and non-invasiveness. However, a major drawback of MRI is the low imaging speed since it needs to perform full data acquisition in *k*-space. To solve this problem, accelerated imaging techniques with under-sampling in *k*-space have been proposed. Among them, parallel imaging (PI)[1] and compressed sensing (CS)[2] are two typical reconstruction methods for acquiring artifact-free images.

Reconstruction methods for PI[1] are divided into image and *k*-space domain algorithms. The sensitivity encoding (SENSE)[3] algorithm removes the aliasing artifacts by solving inverse problems in the image domain. Generalized auto-calibrating partially parallel acquisition (GRAPPA)[4] interpolates non-sampled *k*-space data using AutoCalibration Signals (ACS) sampled in the central *k*-space. To further accelerate parallel imaging, CS reconstruction methods have been proposed. In order to apply the CS theory to MRI reconstruction, a suitable transform domain is needed that makes signal sparse, e.g., wavelet transform[5], total variation (TV)[6], and low rank[7]. Meanwhile, the L1 minimization problem can be solved with regularization terms by using the prior information of the image. However, CS reconstructions are limited by the fact that the under-sampling mask must be incoherent. Moreover, since the solutions of both reconstruction methods require iterative computations, the reconstruction time would be too long. Besides, hand-crafted regularization terms are usually too simple, and several hyper-parameters need to be tuned before application. Thus, conventional CS-MRI technique is limited to acceleration factors of 2~3[8, 9].

It is of note that the advantage of deep learning reconstructions over conventional CS is the greatly reduced reconstruction time while maintaining superior image quality[10, 11]. As for single-channel imaging, Wang et al.[12] developed a convolutional neural network (CNN) to identify the mapping relationship between the zero-filled (ZF) images and the corresponding fully-sampled data. Yang et al.[13] developed a novel deep architecture to include iterative processes in the Alternating Direction Method of Multipliers (ADMM) algorithm into the optimization of a CS-based MRI model. Schlemper et al.[14] used a deep cascade of CNNs to reconstruct under-sampled 2D cardiac MR images. This method outperformed CS approaches in terms of reconstruction error and speed. Yang et al.[15] proposed deep de-aliasing generative adversarial networks (DAGAN) for fast CS-MRI reconstruction. They adopted a U-net architecture as the generator network and coupled an adversarial loss with a novel content loss that could preserve perceptual image details. Quan et al.[10] developed a GAN with a cyclic loss for MRI de-aliasing. This network is composed of two cascaded residual U-Nets, with the first to perform the reconstruction and the second to refine it. Mardani et al.[16] trained a deep residual network with skip connections as a generator with a mixed cost loss of least squares (LS) and L1/L2 norm to reconstruct high-quality MR images. Shaul et al.[17] proposed a two-stage GAN to estimate missing $k$-space samples and to remove aliasing artifacts in the image space simultaneously. Wu et al.[18] integrated the self-attention mechanism into a hierarchical deep residual convolutional neural network (SAT-Net) for improving the sparsely sampled MRI reconstruction. Yuan et al.[19] proposed a network that uses the self-attention mechanism and the relative average discriminator (SARA-GAN), in which half of the input data to the discriminator are true and half are false.

All the approaches above are applicable to single-channel MRI reconstruction. Nevertheless, multi-channel PI is a classic solution of physics-based acceleration. It can not only improve the speed of acquisition but also yield better image quality. In 2018, Hammernik et al.[20] introduced a variational network (VN) to reconstruct complex multi-channel MR data. Aggarwal et al.[21] proposed a model-based deep learning architecture to address the multi-channel MRI reconstruction problem, called MoDL. Zhou et al.[22] combined parallel imaging with CNN, named PI-CNN, for high-quality real-time MRI reconstruction. Wang et al.[23] proposed a multi-channel image reconstruction algorithm based on residual complex convolutional neural networks to accelerate parallel MR imaging (DeepcomplexMRI). Liu et al.[24] developed a novel deep learning-based reconstruction framework called SANTIS for efficient MR image reconstruction. Duan et al.[25] developed a variable splitting network (VS-Net) to effectively achieve a high-quality reconstruction of under-sampled multi-coil MR data.

Lv et al.[26] combined sensitivity encoding and generative adversarial networks for accelerated multi-channel MRI reconstruction, developing SENSE-GAN. Sriram et al.[27] proposed GrappaNet architecture for multi-coil MRI reconstruction. The GrappaNet combined traditional parallel imaging methods with neural networks and trained the model end-to-end. Souza et al.[28] proposed dual-domain cascade U-nets for multi-channel MRI reconstruction. They demonstrated that dual-domain methods are better when simultaneously reconstructing all channels of multi-channel data.

All the above methods need a large number of training samples to train the network parameters and to achieve robust generalization performances. Most previous studies have validated their reconstruction performances on publicly available datasets. However, in clinical applications, it is difficult to obtain tens of thousands of multi-channel data for model training since saving the raw *k*-space data is not included in the routine clinical flow. Thus, it is crucial to improve the generalization of learned image reconstruction networks trained from public datasets. Nowadays, several transfer learning studies have been performed to solve this problem. Han et al.[29] developed a novel deep learning approach with domain adaptation to reconstruct high quality images from under-sampled *k*-space data in MRI. The proposed network employed a pre-trained network using CT datasets or synthetic radial MR data, with fine-tuning using a small number of radial MR datasets. Knoll et al.[30] investigated the effects of image contrast, signal-to-noise-ratio (SNR), sampling pattern, and image content on the generalizability of a pre-trained model and demonstrated the potential for transfer learning with the VN architecture. Dar et al.[31] proposed a transfer-learning approach to examine the generalization capability of networks trained on natural images to T1-weighted and T2-weighted brain images. Arshad et al.[32] evaluated the generalization capability of a trained U-net for the single-channel MRI reconstruction problem with MRI performed using different scanners with different magnetic field strengths, different anatomies, and different under-sampling masks.

However, none of the above-mentioned studies have exploited the generalization capability of multi-channel MRI reconstruction models. In this study, we aimed to examine the generalizability of a pre-trained PI-GAN model for under-sampled MRI

reconstruction under the following situations:

1. Transfer learning of the pre-trained PI-GAN model for a private clinical brain test dataset with tumors.
2. Transfer learning of the pre-trained PI-GAN model for public knee and private liver test datasets.
3. Transfer learning of the pre-trained PI-GAN model under AFs of 2 and 6 for a test brain dataset undersampled by AF of 4.

The implementation of our method is available via GitHub (https://github.com/ljdream0710/TransferLearning_PIGAN).

## 2. Materials and Methods

### 2.1 Problem Formulation

For parallel imaging, the multi-channel image reconstruction problem can be formulated as:

$$y = MFSx + n, \tag{1}$$

in which $M$ is the under-sampling mask; $F$ represents the Fourier transform, $S$ is the coil sensitivity maps, $n$ is the noise, $x$ is the desired image that we want to solve and $y$ represents the acquired $k$-space measurements.

To address the inverse problem of Equation (1), CS-MRI constrains the solution space by introducing some a priori knowledge. Thus, the optimization problem can be expressed as:

$$x = \min_{x} \tfrac{1}{2}\|MFSx - y\|_2^2 + \lambda \Re(x), \tag{2}$$

where the first term represents data fidelity in the $k$-space domain, which guarantees the consistency of reconstruction results with the original under-sampled $k$-space data; $\Re(x)$ denotes the prior regularization term. Term $\lambda$ is a balance parameter, which determines the tradeoff between the prior information and the data fidelity term. In particular, $\Re(x)$ is usually an $L_0$ or $L_1$ norm in a certain sparsity transform domain. To solve the above optimization problem, an iterative method is usually required. With the introduction of deep learning, $\Re(x)$ can be represented by a CNN-based regularization term, that is

$$x = \min_x \frac{1}{2}\|MFSx - y\|_2^2 + \lambda\|x - f_{CNN}(x_u|\theta)\|_2^2, \tag{3}$$

where $f_{CNN}(x_u|\theta)$ is the output of CNN with the parameters $\theta$ and the parameters of the model can be trained with the training dataset. Also, $x_u = F_u^H y$ is the reconstructed ZF images from under-sampled *k*-space data, in which $H$ denotes the Hermitian transpose operation. Recently, conditional GAN is also integrated into MRI reconstruction. GAN consists of a generator *G* and a discriminator *D*. Both the generator and discriminator need to be trained. During the training process, the generator *G* can be trained to estimate the distribution of the given true data and generate data to trick the discriminator *D*. The goal of discriminator *D* is to distinguish the output of the Generator *G* and the true data. Then, after training, we can use the generator alone to generate new samples, which are similar to the actual samples. Therefore, we introduced the conditional GAN loss into the MRI reconstruction, that is

$$\min_{\theta_G} \max_{\theta_D} \mathcal{L}_{cGAN}(\theta_G, \theta_D) =$$

$$\mathbb{E}_{x_t \sim p_{train}(x_t)}[\log D_{\theta_D}(x_t)] + \mathbb{E}_{x_u \sim p_G(x_u)}\left[-\log D_{\theta_D}\left(G_{\theta_G}(x_u)\right)\right], \tag{4}$$

in which $x_u$ is the ZF image which represents the input of the generator, $\hat{x}_u$ is the corresponding reconstructed image yielded from the generator and the fully-sampled ground truth is $x_t$.

## 2.2 PI-GAN Model Architecture

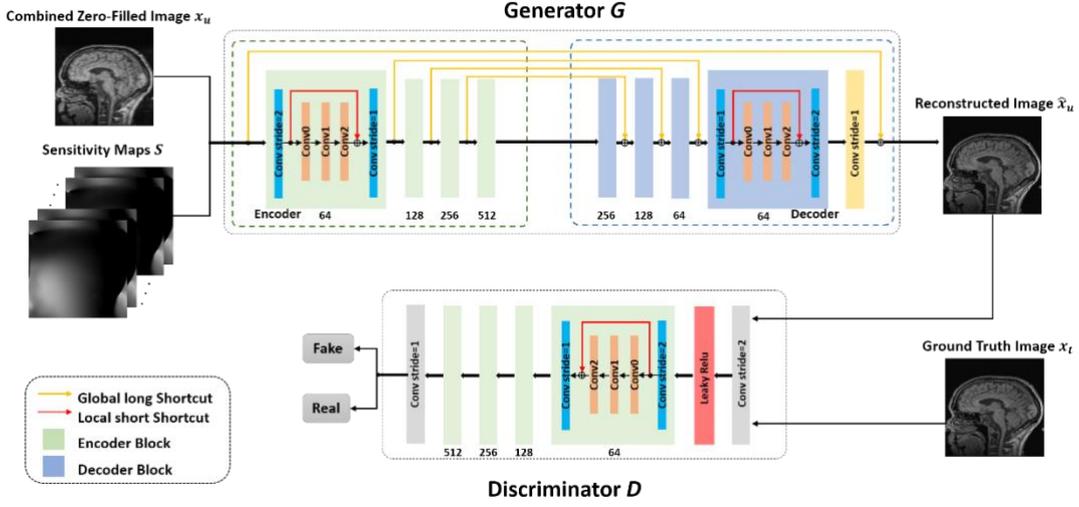

**Fig. 1.** Overview of the PI-GAN architecture. The input of generator $G$ is the combined ZF image $x_u$ and the sensitivity maps $S$ and the output of $G$ is $\hat{x}_u$. $G$ is composed of a residual U-net with 4 encoder (green box) and 4 decoder (lavender box) blocks

Briefly, the PI-GAN architecture integrates data fidelity and regularization terms into the generator. Thus, this approach can not only provide "end-to-end" under-sampled multi-channel image reconstruction but also makes full use of the acquired multi-channel coil information. Besides, to better preserve image details in the reconstruction process, the adversarial loss function is combined with the pixel-wise loss in the image and frequency domains. The loss function is divided into three parts, including the mean absolute error (MAE) in the image domain ($L_{\text{iMAE}}$), and the MAE in the frequency domains $L_{\text{fMAE,M}}$ and $L_{\text{fMAE,1-M}}$, respectively. We choose MAE as the loss function because it provides better convergence than the widely used MSE loss[10, 33, 34]. The input and output of the generator were $x_u$ and $\hat{x}_u$, respectively. The Generator is trained by minimizing the following loss:

$$\mathcal{L}_{\text{TOTAL}} = \mathcal{L}_{\text{GEN}} + \alpha \mathcal{L}_{\text{iMAE}} + \beta \mathcal{L}_{\text{fMAE,M}} + \gamma \mathcal{L}_{\text{fMAE,1-M}}, \quad (5)$$

Here $\alpha, \beta$ and $\gamma$ are the hyperparameters that control the trade-off between each loss term. The four loss terms can be denoted as:

$$\min_{\theta_G} \mathcal{L}_{\text{iMAE}} = \|S\hat{x}_u - x_t\|_1, \quad (6)$$

$$\min_{\theta_G} \mathcal{L}_{\text{fMAE,M}} = \|MFS\hat{x}_u - MFx_t\|_1, \tag{7}$$

$$\min_{\theta_G} \mathcal{L}_{\text{fMAE,1-M}} = \|(1-M)FS\hat{x}_u - (1-M)Fx_t\|_1. \tag{8}$$

$$\min_{\theta_G} \mathcal{L}_{\text{GEN}}(\theta_G) = -\log\left(D_{\theta_D}\left(G_{\theta_G}(x_u)\right)\right) \tag{9}$$

GAN is difficult to be trained successfully because of the need for alternate training. Thus, to stabilize the training of GAN, we introduce the refinement learning, which is $\hat{x}_u = G_{\theta_G}(x_u) + x_u$. That means the generator only needs to learn part of data that is not acquired which can significantly reduce the model complexity.

**2.2.1 Generator Architecture**

As shown in **Fig. 1**, the generator architecture of PI-GAN is composed of a residual U-net with 4 encoder (green box) and 4 decoder (lavender box) blocks. The number of feature maps is listed under each block. Besides, each module takes a 4D tensor as input and uses 2D convolution with a filter size of 3×3 and a stride of 2. A global long-skip connection between the corresponding encoder and decoder blocks is used to extract details from the image. The encoder block adopts the down-sampling method. Each encoder block has three parts. The first and third parts are 2D convolutional layers with strides of 2 and 1, respectively; the second part is a residual block with a short-skip connection, including conv0, conv1, and conv2. Among them, conv0 and conv2 are convolutional layers with a stride of 1 and a filter size of 3×3; the number of feature maps is 64. Conv1 is a convolutional layer with a stride of 1, a filter size of 3×3; the number of feature maps is 32. The decoding block adopts the up-sampling strategy, and each decoder block also consists of three parts. After the last decoding block, a 2D convolutional layer with a filter size of 3×3 and a stride of 1 provides the global information of the feature map.

**2.2.2 Discriminator Architecture**

The discriminator model is a 7-layer CNN network. The first and second layers are convolutional layers with a filter size of 4×4 using the Leaky ReLU as the activation function. The third to sixth layers use the same network structure as the generator encoder block. The last layer is a convolutional layer with a stride of 1.

**2.3 Datasets**

2.3.1 Public MRI dataset

Datasets of healthy subjects were obtained from the Calgary-Campinas brain MR raw data repository (https://sites.google.com/view/calgary-campinas-dataset/home) for training. The raw data were acquired using the 3D T1 gradient-echo (GRE) sequence on a clinical MR scanner (Discovery MR750; General Electric Healthcare, Waukesha, WI) with a 12-channel head-neck coil. The acquisition parameters are: Repetition Time (TR)/Echo Time (TE)=6.3 ms/2.6 ms; Inversion Time (TI)=650 ms; 170 contiguous 1.0 mm slices; the field of view (FOV), 256×218 mm$^2$.

The knee datasets used in this study were from the "Stanford Fully Sampled 3D FSE Knees" repository (http://old.mridata.org/fullysampled/knees). The raw data were acquired using the 3D TSE sequence with proton density weighting including fat saturation comparison on a 3.0T whole body MR system (Discovery MR 750, DV22.0, GE Healthcare, Milwaukee, WI, USA) with 8 coils. The acquisition parameters were: TR/TE=1550 ms/25 ms; slice thickness, 0.6 mm; 256 slices; FOV=160×160 mm$^2$; the size of the acquisition matrix, 320×320. The voxel size was 0.5 mm. We randomly selected 18 subjects for network tuning and 2 subjects for testing, corresponding to 1800 and 200 2D images, respectively.

2.3.2 Private MRI dataset

All subjects gave their informed consent for inclusion before they participated in the study with approval from the local institutional review board (in accordance with the Declaration of Helsinki). The institutional review board (at Shanghai Ruijin Hospital) has approved the MRI scanning.

Private brain tumor MRI datasets were acquired in 17 subjects using three different imaging sequences on the 3T Philips Ingenia MRI system (Philips Healthcare, Best, the

Netherlands) scanner, including T1-weighted imaging (T1W), T1-weighted sagittal imaging (T1SAG), and FLAIR imaging, with an 8-channel head-neck coil. T1W images were acquired using the inversion recovery (IR) sequence with the following parameters: TR/TE, 1800 ms/20 ms; slice thickness, 5 mm. The matrix size of the data was 480×480×20. FLAIR images were acquired using spectral pre-saturation by the inversion recovery (SPIR) sequence with the following parameters: TR/TE, 7000 ms/120 ms, slice thickness, 5 mm. The matrix size of the data was 384×384×20. We randomly selected 10 subjects for network tuning and 7 for testing, corresponding to 216 and 89 images, respectively. Meanwhile, T1SAG images were acquired using the fast field echo (FFE) sequence with the following parameters: TR/TE, 250 ms/2.3 ms; slice thickness, 5 mm. The matrix size of the data was 512×512×20. We randomly selected 10 subjects for network tuning and 7 subjects for testing, corresponding to 264 and 75 images, respectively.

Private liver MRI datasets were acquired in 16 subjects. The images were fully-sampled using the T2-weighted turbo spin-echo pulse sequence on Philips 3T MRI scanner using an 8-channel coil. The parameters are as follows: TR/TE, 2750 ms/78 ms; slice thickness, 6 mm. The matrix size of the data was 430×430×40. We randomly selected 14 subjects for network tuning and 2 for testing, corresponding to 326 and 66 2D images, respectively.

**2.4 Transfer Learning**

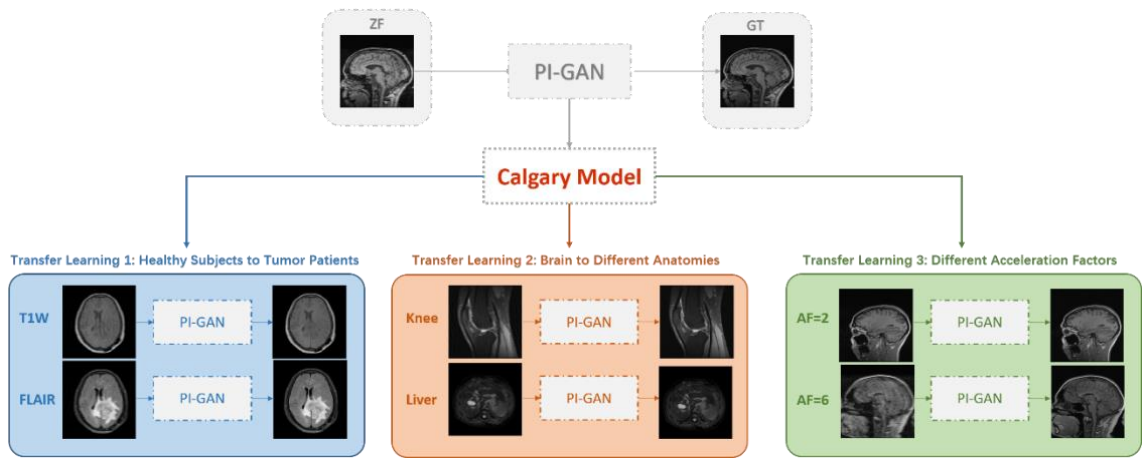

**Fig. 2. Transfer learning of a pre-trained PI-GAN model for under-sampled MRI reconstruction.**

**The Calgary Model was initially trained on healthy brain images (AF=4).**

As shown in **Fig.2**, we utilized transfer learning of a pre-trained PI-GAN model to reconstruct the under-sampled MRI data for three different scenarios. In the pre-training step, we used Calgary-Campinas raw brain data to train our PI-GAN model. The model is named as the "Calgary Model". The dataset included 4700 images from 47 subjects for training and 2000 images from 20 subjects for testing. During the training process, we separated the dataset into training and validation. During each iteration, 20 images were randomly selected for validation. Each model with the highest Peak Signal to Noise Ratio (PSNR), i.e., with the best performance, in the validation dataset was selected for further independent testing. All fully sampled *k*-space data were retrospectively under-sampled via random sampling trajectories for AF=4 with 24 fixed central lines (ACS = 24). We trained the networks with the following hyperparameters: $\alpha = 1$ and $\beta = \gamma = 10$ with Adam[35] optimizer. Essentially, the selection of hyperparameters is based on the previous work[10] The selection rules are the magnitudes of different loss terms that should be balanced into the same scale. The model used a batch size of 8 and the initial learning rate of $10^{-4}$ for training, which decreased monotonically over 1000 epochs. The weights obtained from the pre-trained model were then used as initial weights in the other transfer learning processes.

The sensitivity maps of each coil were estimated using ESPIRiT[36]. The experiments were implemented in Python3 using Tensorflow (https://www.tensorflow.org/) as the backend. The reconstruction processes were carried out on a workstation using specifications as Intel(R) Xeon(R) CPU E5-2698 v4 @ 2.20GHz with 256 GB RAM and an NVIDIA GV100GL (Tesla V100 DGXS 32GB) graphics processing unit (GPU).

In order to evaluate the transfer learning performances, we compared the obtained images with those reconstructed by other methods. The L1-ESPIRiT reconstruction was performed using the Berkeley Advanced Reconstruction Toolbox (BART)[37]. The "Directly Trained" model stands for the model trained with the current dataset including limited image numbers. The "Calgary model" is trained on the Calgary dataset without tuning. The "Transfer Learning" model fine-tunes the pre-trained Calgary Model for the reconstruction with the current dataset. **Table 1** shows the number of images used for training and test among different models.

**Table 1. Number of images used for training and test.**

| Model Name | Fine-tuning | Training Dataset | | | | | Test Dataset | | | | |
|---|---|---|---|---|---|---|---|---|---|---|---|
| | | Brain | | | Knee | Liver | Brain | | | Knee | Liver |
| | | T1W | FLAIR | T1SAG | | | T1W | FLAIR | T1SAG | | |
| Directly Trained | × | 216 | 216 | 264 | 1800 | 326 | 89 | 89 | 75 | 200 | 66 |
| Transfer Learning | ✓ | | | | | | | | | | |
| Calgary Model | × | 4700 T1W images of Calgary-Campinas | | | | | | | | | |

## 2.5 Quantitative Evaluation

The obtained reconstruction results were evaluated using three metrics: peak signal to noise ratio (PSNR), structural similarity index measure (SSIM) and normalized root mean square error (RMSE). Besides, histograms of reconstruction results were generated from a region of interest (ROI) covering the brain tumor, and histogram parameters (kurtosis [k] and skewness [s]) were obtained.

## 3. Results

The "Calgary Model" model was pre-trained using a large number of images from the public Calgary dataset. This model achieved satisfactory reconstruction results (PSNR, 32.60±0.89; SSIM, 0.90±0.01; RMSE, 0.024±0.002). Then, we transferred this model to the test dataset from different domains, which had only a few hundred images.

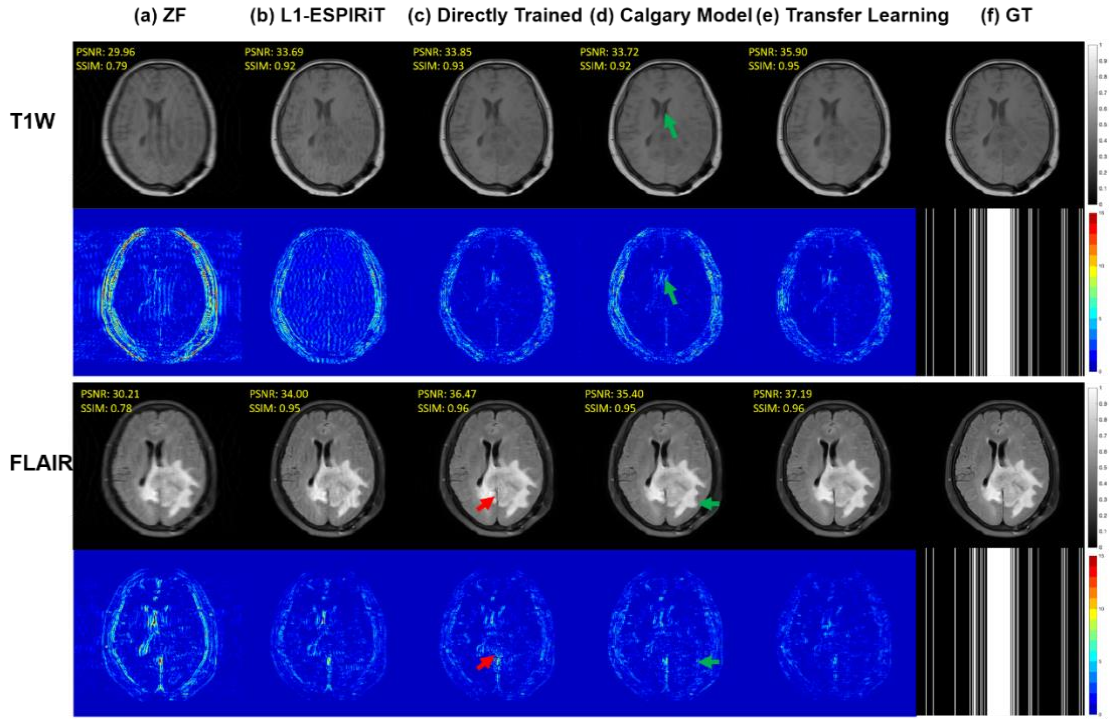

**Fig. 3.** Typical reconstructed T1W and FLAIR images at the same slice from the brain tumor dataset. (a) ZF image, (b) L1-ESPIRiT, (c) Directly Trained, (d) Calgary Model (e) Transfer learning, and (f) GT. The second and fourth rows show error maps for various reconstruction images and the under-sampling mask with AF=4.

**Fig. 3** shows the test results of different reconstruction methods on T1W and FLAIR images. As shown in T1W images, the L1-ESPIRiT reconstruction results had significant blurring. As indicated by green arrowheads, the results of Directly Trained (PSNR, 33.85; SSIM, 0.93) were slighter better than those of Calgary Model (PSNR, 33.72; SSIM, 0.92), which contained residual artifacts. The Transfer Learning reconstruction (PSNR, 35.90; SSIM, 0.9) method outperformed other reconstruction methods. In the corresponding FLAIR images, L1-ESPIRiT still had some blurring. In the tumor region (red arrowheads), the Directly Trained image (PSNR, 36.47; SSIM, 0.96) produced some artifacts that were not eliminated due to a small training dataset. As shown by the green arrow, the Calgary Model image (PSNR, 35.40; SSIM, 0.95) still contained residual artifacts, which were worse than in the Transfer Learning image (PSNR, 37.19; SSIM, 0.96).

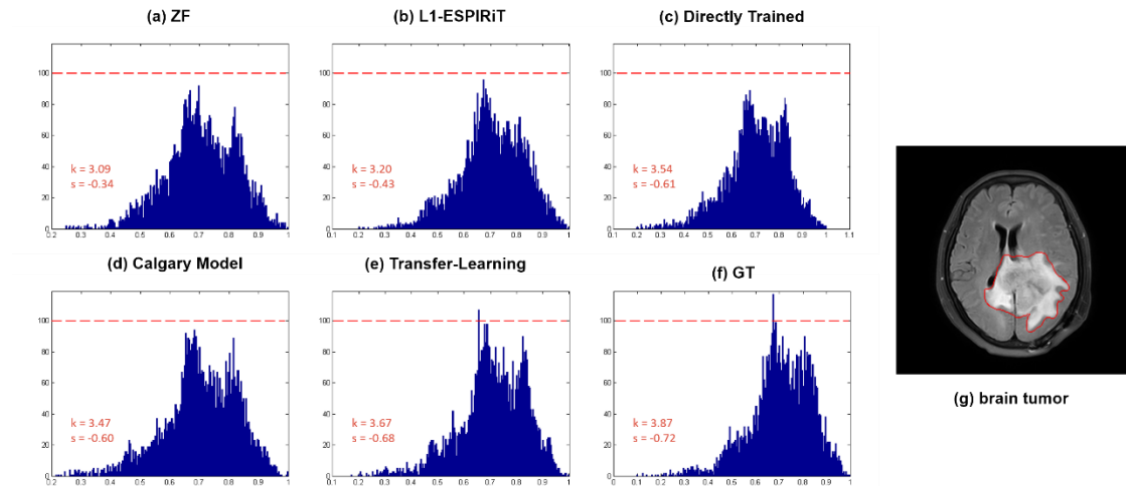

**Fig. 4** Histograms analysis of the tumor in the corresponding slices in Fig.3 were performed by calculating the data within the ROI marked with red contour (g) The data are normalized to [0, 1] for convenient display. The kurtosis (notated as k) and skewness(notated as s) are also calculated.

**Fig. 4** shows histograms of the tumors in the corresponding slices in Fig.3. The histogram and quantitative indicators show that the Transfer Learning model (k=3.67, s=-0.68) achieved the closest quantifications to GT (k=3.87,s=-0.72) as compared to other methods.

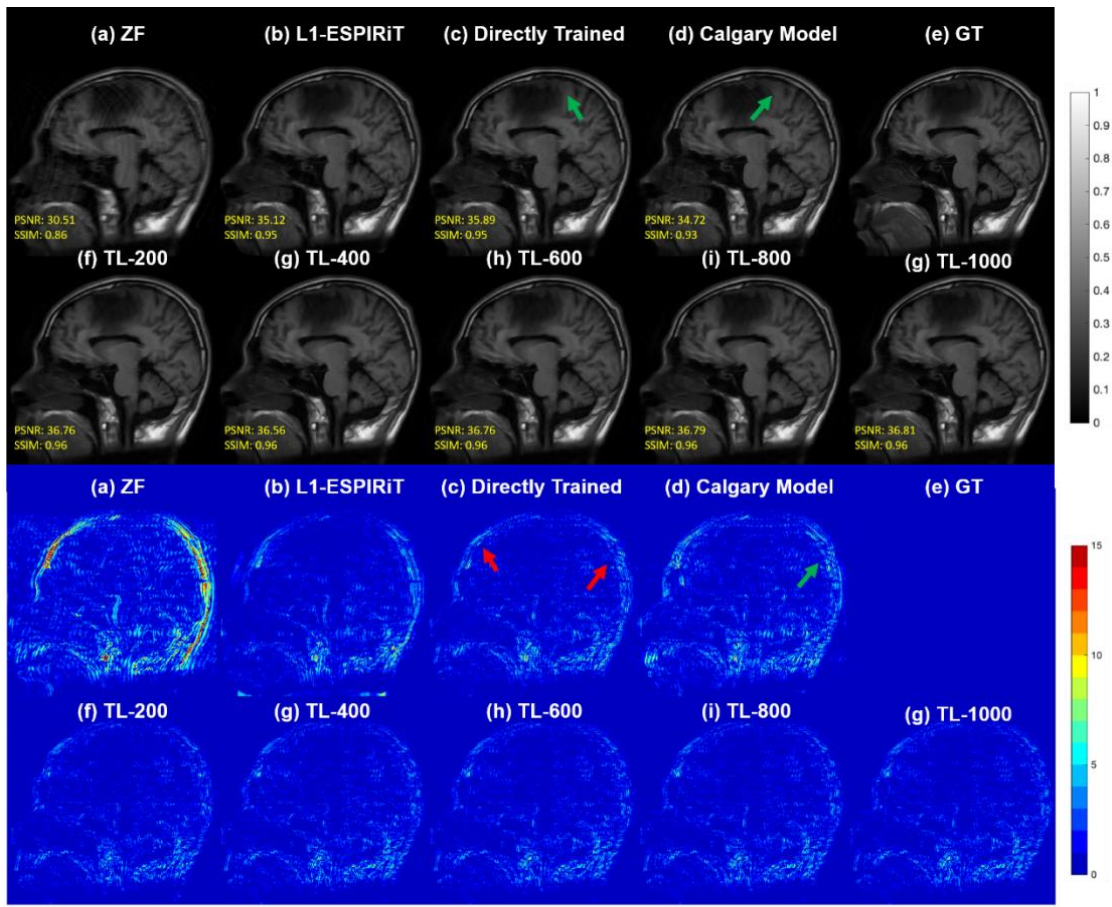

**Fig. 5.** Representative reconstructed sagittal T1W brain tumor images at AF=4. (a) ZF image, (b) L1-ESPIRiT, (c) Directly Trained, (d) Calgary Model, (e) GT, and (f-g) Transfer Learning (Fine-tuning of the pre-trained PI-GAN using 200~1000 epoch). The third and fourth rows show error maps for each reconstruction image.

**Fig. 5** shows the test results of different reconstruction methods and transfer learning with the Calgary model over 200~1000 epochs on sagittal T1W images. L1-ESPIRiT data were still contaminated with blurring at brain edges. As shown by green arrowheads, both Directly Trained and Calgary Model contained residual artifacts. However, as indicated by green arrowheads in error maps, the edges of the Calgary Model images were very noisy, while the Directly Trained image was free of artifacts at the corresponding site. This indicated that Directly Trained (PSNR, 35.89; SSIM, 0.95) images were better than those of the Calgary Model (PSNR, 34.72; SSIM, 0.93) group. As indicated by red arrowheads, Transfer Learning data could remove the artifacts found in Directly Trained images, and the brain edges were smoother. We used a total of 1000 epochs for fine-tuning, and the results showed that TL200 (PSNR, 36.76;

SSIM, 0.96) had almost reached the optimum, which may be due to fast convergence because training and test data were both from the brain.

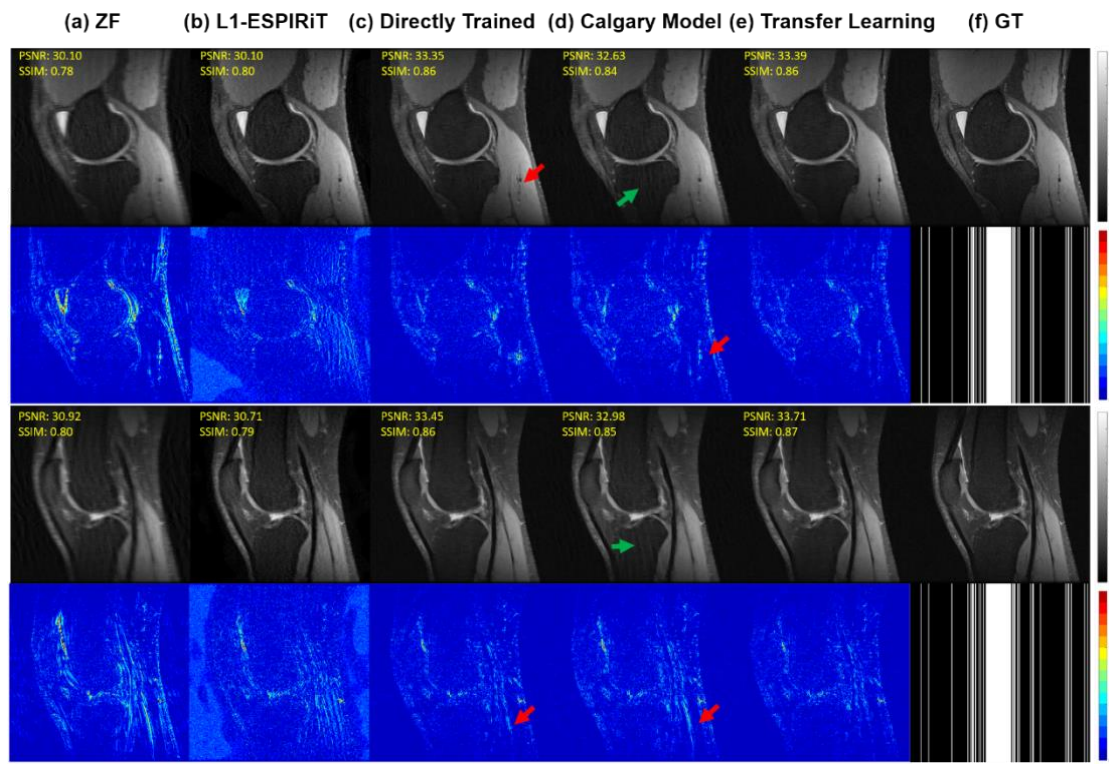

**Fig. 6. Typical reconstructed knee images at AF=4. (a) ZF image, (b) L1-ESPIRiT, (c) Directly Trained, (d) Calgary Model, (e) Transfer learning, and (f) GT. The second and fourth rows show error maps for each reconstructed image and the under-sampling mask with AF=4.**

**Fig. 6** shows the results of the different reconstruction approaches tested on Knee images. The results showed that L1-ESPIRiT did not remove the artifacts completely. As indicated by red arrowheads, vessels were blurry in Directly Trained and Calgary Model images but were preserved well in Transfer learning data. Calgary Model data were worse than Directly Trained results because of residual artifacts (shown by green arrowheads). This indicated that it may be better to train an independent model for Knee data with only a small amount of data. It is obvious that Transfer learning yielded the best data among all methods.

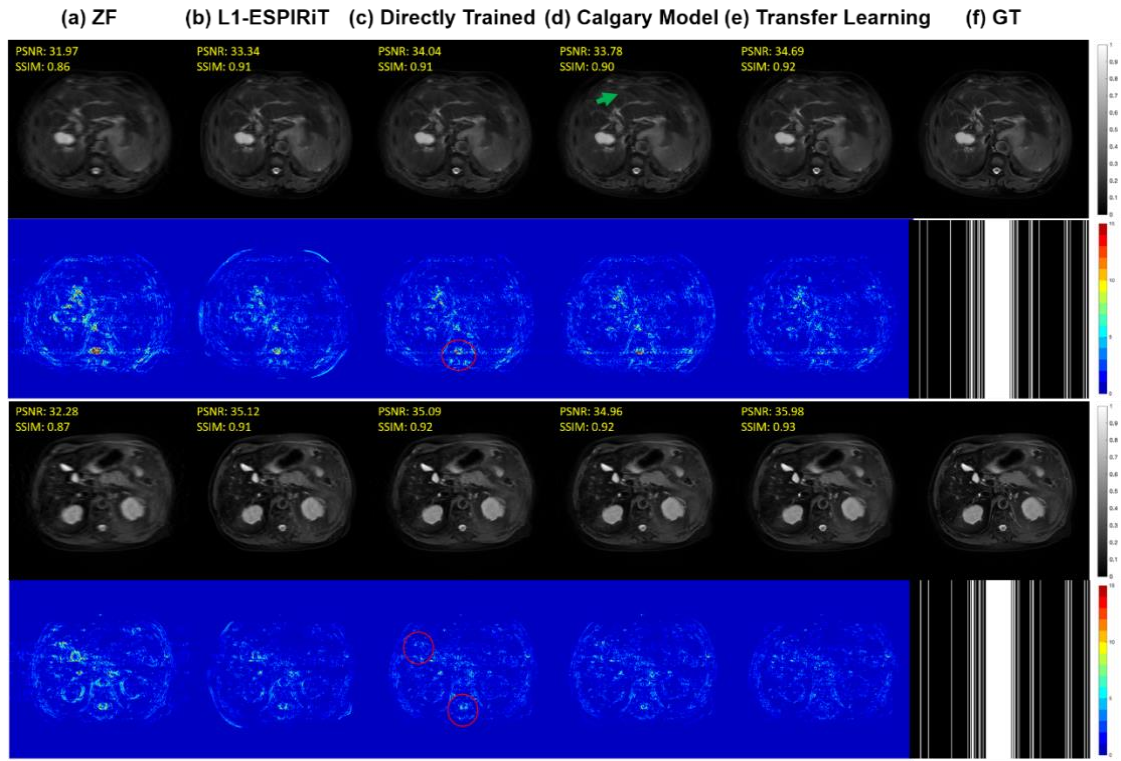

**Fig. 7. Representative reconstructed liver images at AF=4. (a) ZF image, (b) L1-ESPIRiT, (c) Directly Trained, (d) Calgary Model, (e) Transfer learning, and (f) GT. The second and fourth rows show error maps for each reconstructed image and the under-sampling mask with AF=4.**

**Fig. 7** shows the reconstruction results obtained with different methods on Liver tumor images. As shown in the figure, L1-ESPIRiT reconstruction had image blurring. Besides, Calgary Model data contained obvious residual artifacts (indicated by green arrowheads). The SSIM value of Calgary Model data was 0.01 lower than that of Directly Trained images. Besides, the vessels of Directly Trained were not sharp enough as shown by red circles, and errors were much larger. The Transfer learning shows the best results.

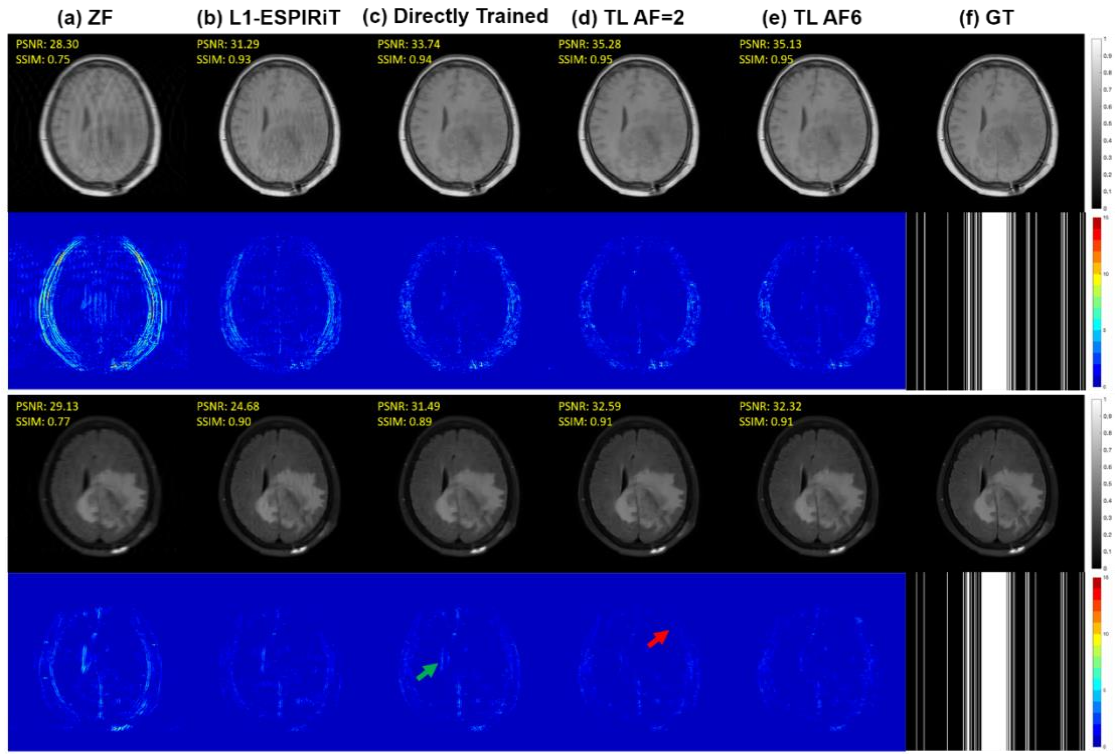

**Fig. 8. Representative reconstructed T1W and FLAIR brain tumor images using different models. (a) ZF image, (b) L1-ESPIRiT, (c) Directly Trained, (d) TL AF=2 (Transfer Learning of the model pre-trained on Calgary dataset with AF=2), (e) TL AF=6 (Transfer Learning of the model pre-trained on Calgary dataset with AF=6) and (f) GT. The second and fourth rows show error maps for each reconstruction image and the under-sampling mask with AF=4.**

**Fig. 8** shows the test results of different reconstruction methods in brain tumor datasets including T1W and FLAIR images. L1-ESPIRiT has the worst reconstruction method. Directly Trained images contained residual artifacts as shown by green arrowheads. After transfer learning of the model pre-trained on the Calgary dataset with AF=2 or AF=6, reconstruction results were significantly improved. As shown by red arrowheads, transfer learning with AF=2 had much better quantitative values and sharper edge reconstruction.

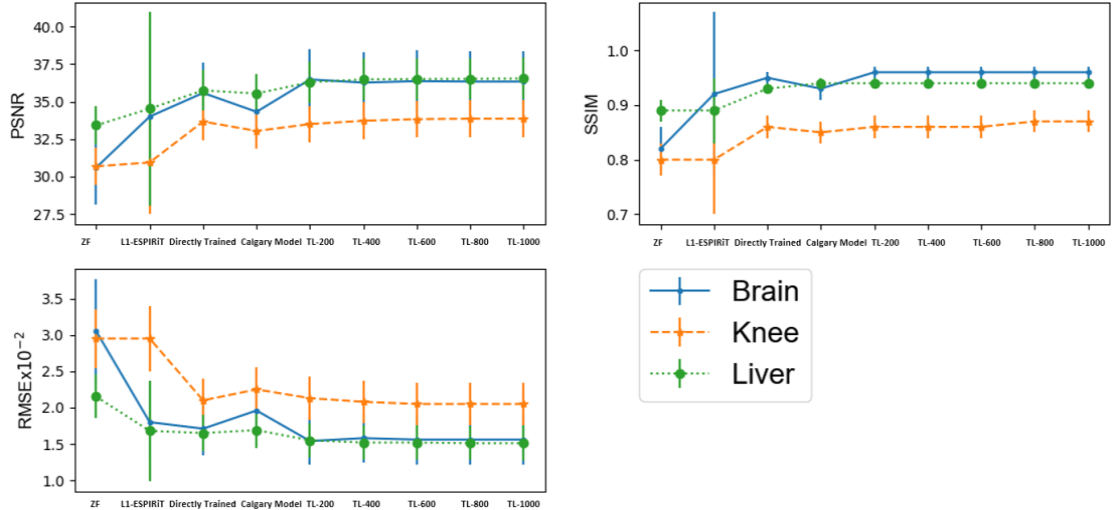

**Fig. 9. Reconstruction performances for the under-sampled brain, knee, and liver images with AF=4. The average and standard deviation of PSNR, SSIM, and RMSE (×10⁻²) values across test images were measured for ZF, L1-ESPIRiT, Directly Trained, Calgary Model, and Transfer Learning (TL) with 200~1000 epochs.**

**Fig. 9** shows PSNR, SSIM, and RMSE ($\times 10^{-2}$) values for the images reconstructed with different reconstruction models at different anatomical areas under 4× acceleration. All the results showed that Transfer Learning was better than that of Calgary Model. Meanwhile, the brain results (blue line) showed that the TL200 (PSNR, 36.48±2.02; SSIM, 0.96±0.01; RMSE, 1.54±0.33) index values were optimal, indicating that the model converged quickly after a simple fine-tuning of the brain dataset. Knee (orange dashed line) and Liver (green dotted line) data showed that convergence was reached with TL800 for knee (PSNR, 33.86±1.24; SSIM, 0.87±0.02; RMSE, 2.05±0.29) and liver (PSNR, 36.54±1.39; SSIM, 0.94±0.01; RMSE, 1.51±0.24) images, which was low but still better than that obtained by training directly with knee or liver data.

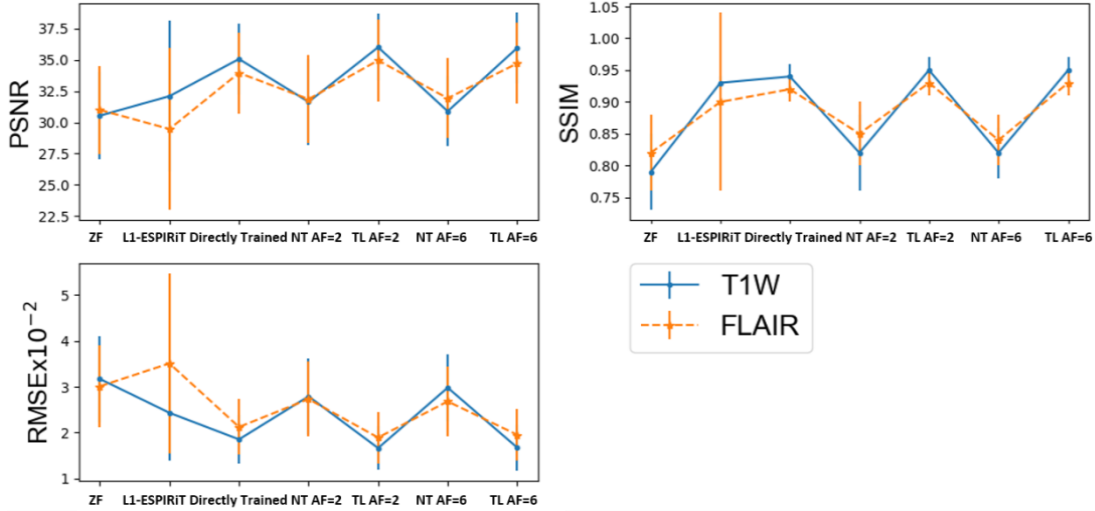

**Fig. 10.** PSNR, SSIM, and RMSE (×10⁻²) values of T1W and FLAIR images reconstructed using different models under AF=4. NT AF=2, testing the model trained using Calgary dataset under-sampled by AF=2 directly. TL AF=2, using transfer learning to fine tune the model trained by the Calgary dataset with AF=2. NT AF=6, testing the model trained using Calgary dataset under-sampled by AF=6 directly. TL AF=6, using transfer learning to fine tune the model trained by the Calgary dataset with AF=6.

**Fig. 10** shows reconstruction results for T1W and FLAIR images fine-tuned with different acceleration factors. Directly Trained results showed better results than NT AF=2 and NT AF=6. In addition, by performing fine-tuning, both TL AF=2 for T1W (PSNR, 36.00±2.67; SSIM, 0.95±0.02; RMSE, 1.66±0.48) and FLAIR (PSNR, 34.95±3.26; SSIM, 0.93±0.02; RMSE, 1.89±0.56) images and TL AF=6 for T1W (PSNR, 35.95±2.80; SSIM, 0.95±0.02; RMSE, 1.67±0.50) and FLAIR images (PSNR, 34.71±3.24; SSIM, 0.93±0.02; RMSE, 1.95±0.57) images outperformed Directly Trained reconstruction data for T1W (PSNR, 35.06 ± 2.82; SSIM, 0.94±0.02; RMSE, 1.85±0.54) and FLAIR (PSNR, 33.95±3.22; SSIM, 0.92±0.03; RMSE, 2.12±0.61) images. However, it is clear that both T1W and FLAIR data showed the best fine-tuning results for AF=2.

**Table 2. Quantitative assessment of PSNR, SSIM, and RMSE values (mean±standard deviation) across the T1W and FLAIR test images obtained with different reconstruction models.**

|  | T1W | | | FLAIR | | |
| --- | --- | --- | --- | --- | --- | --- |
|  | PSNR | SSIM | RMSE×10$^{-2}$ | PSNR | SSIM | RMSE×10$^{-2}$ |
| ZF | 30.53±3.47 | 0.79±0.06 | 3.17±0.93 | 30.99±3.50 | 0.82±0.06 | 3.01±0.90 |
| L1-ESPIRiT | 32.09±6.02 | 0.93±0.01 | 2.43±1.04 | 29.46±6.48 | 0.90±0.14 | 3.51±1.97 |
| Directly Trained | 35.06±2.82 | 0.94±0.02 | 1.85±0.54 | 33.95±3.22 | 0.92±0.03 | 2.12±0.61 |
| Calgary Model | 34.39±2.58 | 0.93±0.02 | 1.98±0.52 | 33.64±3.11 | 0.91±0.03 | 2.19±0.61 |
| **Transfer Learning** | **35.49±2.79** | **0.95±0.05** | **1.76±0.52** | **34.99±3.29** | **0.93±0.02** | **1.89±0.56** |

As shown in **Table 2**, the Directly Trained model of FLAIR images improved the PSNR by 3.06% and decreased the RMSE by 10.85% after transfer learning compared with ZF reconstruction. Meanwhile, the Directly Trained model of T1W images improved the PSNR by only 1.23%, and the RMSE only decreased by 4.86% compared with ZF reconstruction. However, the values obtained for T1W images after transfer learning outperformed those of FLAIR images. Meanwhile, the PSNR values listed in **Table 2**, suggesting that the Transfer Learning achieved significantly better performance ($p$ values: Directly Trained vs Transfer Learning: 2.64e-16, Calgary Model vs Transfer Learning: 8.30e-16 for T1W and Directly Trained vs Transfer Learning: 2.55e-16, Calgary Model vs Transfer Learning: 2.55e-16 for FLAIR, calculated with the Wilcoxon signed-rank sum test where $p < 0.05$ was considered to be statistically significant difference). Besides, to show more clearly the difference among Directly Trained, Calgary Model and Transfer Learning, we calculated the PSNR, SSIM and RMSE of each test image in Fig. 11. As shown in Fig. 11, all the quantitative results showed that Transfer Learning has the best performance for both T1W and FLAIR images.

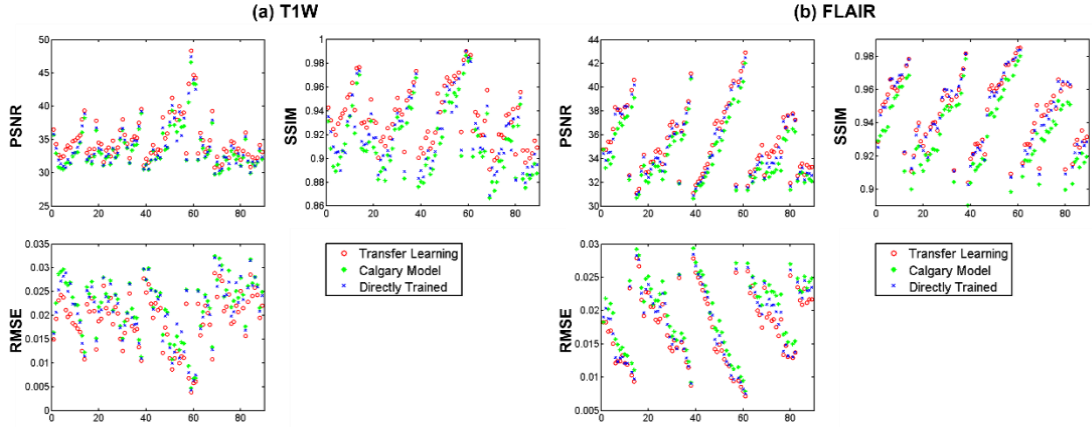

Fig. 11. The PSNR, SSIM, and RMSE values of each test image for different models. For both T1W and FLAIR images, all the quantitative results showed that Transfer Learning has the best performance.

## 4. Discussion

The main contribution of this study was to develop a transfer learning enhanced GAN approach to reconstruct several unseen multi-channel MR datasets. The results demonstrated that transfer learning from a pre-trained model could reduce the variation between the training and test datasets in terms of the variations in image contrast, anatomy and AF.

For the brain tumor dataset, T1W reconstruction images showed better performance compared with FLAIR images. This suggests that the optimal strategy may be to make training and test data with the same contrasts. This is because the Calgary model was initially trained with T1W data. When the training and the test dataset belong to similar distribution, the reconstruction performance will be good. Similarly, the larger the differences between the distributions of training and test datasets, the worse the reconstruction performance will be. Besides, after transfer learning, FLAIR images had much more improved PSNR and SSIM and reduced RMSE than T1W images. This indicates that fine-tuning is more effective for data reconstructed across domains due to the additional information provided by T1W data for FLAIR images. Meanwhile, the above quantitative results support the notion that transfer learning could address the deviation between healthy and diseased subjects. Subhas et al.[38] reported that training a model using images with or without pathology does not affect performance. These observations can be explained that, under the condition of using the same under-sampling trajectory, the PI-GAN model is used for artifact removal and is not related to the image content. However, the present study differs from theirs in that they only included data with pathologies in the training dataset and did not perform fine-tuning of the network model. Moreover, the model trained was only applicable for combined single-channel images in the latter study. Besides, histogram analysis is linked to the application for the tumor heterogeneity assessment[39], which is important for treatment planning [40]. Our results show that the distribution of the reconstructed image using transfer learning is closer to that of the fully sampled image as compared to other methods, which can further facilitate the segmentation and diagnosis of tumor malignancy.

Besides, we successfully transferred the model pre-trained on Calgary data to different anatomies. We found that compared with the knee and liver datasets, brain tumor samples converged faster. This may be because brain tumor data were in the same anatomical position as the training data, so a small number of transfer learning steps could achieve optimal results. Arshad et al.[32] also transferred the U-net model pre-trained on brain samples to heart data. On one hand, the latter study was only applicable to single channel images. On the other hand, it explored the reconstruction performance of the model after fine-tuning using different datasets. In contrast, we explored the performance of model reconstructed images after fine-tuning with a fixed training set and a different number of iterations. It is reasonable that they concluded that the larger the dataset, the higher the performance. We believe that the present investigation is more realistic because the amount of data that can be collected (e.g., liver, kidney, and heart) is inherently small. Results showed that, as long as fine-tuning is performed, whether the acceleration factor is higher or lower than its own under-sampling acceleration factor, reconstruction results would be better than taking only a small amount of its own data for training. However, it is clear that both T1W and FLAIR data showed that fine-tuning was best for models with AF=2, which means that a model with a low acceleration factor should be selected for transfer learning.

This study had some limitations. First, we under-sampled the used MRI data retrospectively, which might not be as good as prospective data under-sampling of *k*-space. Additional studies using prospective under-sampling are needed to validate these results. Secondly, several unsupervised learning algorithms[41-43] have been proposed to address the problem of insufficient sample size. In the future, we will compare the reconstruction performance of our transfer learning method with the existing unsupervised learning strategies. Finally, the PI-GAN model requires accurate sensitivity maps during the training process. However, we have not explicitly validated the quality of the sensitivity maps before performing the reconstruction. The optimization of the sensitivity maps has been done by other studies[44, 45] , which is beyond the scope of our study.

## 5. Conclusion

This study provides insights into the generalization ability of a learned PI-GAN model for under-sampled multi-channel MR images with respect to deviations between the training and test datasets. Our results indicate that the PI-GAN model pre-trained on public Calgary brain images can be applied to brain tumor patients with T1W and FLAIR images, knee and liver images, and images with different acceleration factors through transfer learning with a small tuning dataset. Thus, this study reveals the potential of transfer learning in multi-channel MRI reconstruction, where no sufficient data are available for complete training.

## 6. Acknowledgements

This research was supported by the National Natural Science Foundation of China (No. 61902338, No. 62001120), the Shanghai Sailing Program (No. 20YF1402400), the European Research Council Innovative Medicines Initiative on Development of Therapeutics and Diagnostics Combatting Coronavirus Infections Award 'DRAGON: rapiD and secuRe AI imaging based diaGnosis, stratification, fOllow-up, and preparedness for coronavirus paNdemics [H2020-JTI-IMI2 101005122], the British Heart Foundation [PG/16/78/32402], the AI for Health Imaging Award 'CHAIMELEON: Accelerating the Lab to Market Transition of AI Tools for Cancer Management' [H2020-SC1-FA-DTS-2019-1 952172], and the Hangzhou Economic and Technological Development Area Strategical Grant [Imperial Institute of Advanced Technology].